\documentstyle[graphicx,multicol,eqsecnum,prb,aps]{revtex}

\begin{document}
\draft
\preprint{HEP/123-qed}
\title {Spin-phonon coupled modes in the incommensurate phases of doped CuGeO$_{3}$}
\author {K. Takehana, T. Takamasu, M. Hase, and G. Kido}
\address {Physical Properties Division, National Research Institute for Metals,\\
3-13 Sakura, Tsukuba, Ibaraki 305-0003, Japan}
\author{T. Masuda and K. Uchinokura}
\address {Department of Advanced Materials Science, The University of Tokyo,\\ 6th Engineering Building, 7-3-1 Hongo, Bunkyo-ku, Tokyo 113-8656, Japan}
\date{\today}
\begin{multicols}{2}[
\maketitle
\begin{abstract}
The doping effect of the folded phonon mode at 98 cm$^{-1}$ was investigated on the Si-doped CuGeO$_3$ by magneto-optical measurements in far-infrared (FIR) region under high magnetic field. 
The folded phonon mode at 98 cm$^{-1}$ appears not only in the dimerized (D) phase but also in the dimerized-anitiferromagnetic (DAF) phase on the doped CuGeO$_3$. 
The splitting was observed in the incommensurate (IC) phase and the antiferromagnetically ordered incommensurate (IAF) phase above $H_C$. 
The split-off branches exhibit different field dependence from that of the pure CuGeO$_3$ in the vicinity of $H_C$, and the discrepancy in the IAF phase is larger than that in the IC phase. 
It is caused by the interaction between the solitons and the impurities.
\end{abstract}
\pacs{78.66.-w,78.30.-j,63.22.+m}
]
\narrowtext

\section{Introduction}
\label{sec:level1}

An inorganic compound CuGeO$_{3}$ was discovered to undergo the spin-Peierls (SP) transition at the critical temperature $T_{SP}$ = 14 K in magnetic susceptibility measurements.\cite{Hase} 
In the SP transition, the formation of the superlattice occurs owing to the coupling between the one-dimensional spin system and three-dimensional phonon systems, 
together with the opening of a magnetic energy gap between the singlet ground state and the triplet excited state. 
The lattice dimerization was confirmed by electron diffraction,\cite{Kaminura} x-ray and neutron diffraction.\cite{Pouget,Hirota} 
The opening of the spin gap due to the SP transition was directly observed by inelastic neutron scattering (INS) experiments.\cite{Nishi} 
A phase transition from the dimerized (D) phase to the incommensurate (IC) phase was found at the critical field $H_C \approx$ 12 T.\cite{Hase2} 
The incommensurate lattice modulation was observed in x-ray measurements under high magnetic field above $H_C$.\cite{Kiryukhin1} 
Higher order harmonics of the incommensurate Bragg reflections were also 
observed just above $H_C$, which indicates that the lattice 
modulation in the IC phase forms a soliton lattice.\cite{Kiryukhin2} 

One of the most important studies that the discovery of CuGeO$_3$ enables is that of impurity effects on the quasi-one-dimensional spin system. 
For small amounts of impurity, another phase transition to the dimerized-antiferromagnetic (DAF) phase was found at $T_N$ which is lower than $T_{SP}$.\cite{Hase3} 
In the DAF phase, the coexistence of the lattice distortion and the antiferromagnetic ordering was found in neutron diffraction experiments.\cite{Regnault,Sasago,Martin} 
Recently, Masuda {\em et al.} found a first-order compositional phase transition between the DAF and the uniform-antiferromagnetic (UAF) phases in Mg-doped CuGeO$_3$.\cite{Masuda1} 
It is not clear whether this kind of the phase transition exists in 
Si-doped CuGeO$_3$ or not but the sample with low concentration used in this paper has 
apparently the nature of DAF phase below $T_N$.\cite{Masuda2}
There are some reports about the nature of the high-field phase in doped CuGeO$_3$ with low concentration of impurities. 
Namely, two phase transitions have been reported in specific heat experiments: one was the phase transition between uniform (U) and IC phases, 
and the other was a transition between the IC phase and the antiferromagnetically ordered incommensurate (IAF) phase.\cite{Hiroi} 
On the other hand, ultrasonic velocity measurements have suggested only one phase transition occurred above $H_C$.\cite{Poirier}
Recently, the different behavior between the IC and IAF was reported on the thermal conductivity.\cite{Takeya} 
The authors suggested that it is caused by the ^^ ^^ freezing" of the solitons in the IAF phase. 

Optical studies are quite sensitive for investigation of the change in crystal structure that is caused by a phase transition. 
Factor group analysis predicts that nine infrared-active folded modes and 18 Raman-active folded modes appear below $T_{SP}$ in addition to optical phonons in the U phase, 
which occur when the symmetry of the crystal structure is lowered from Pbmm of the U phase\cite{Vollenkle} to  Bbcm of the D phase.\cite{Hirota,Braden} 
Three Raman modes at 107, 369 and 820  cm$^{-1}$ were assigned to the A$_{g}$ folded phonons in the D phase,\cite{Kuroe1} with the first one having a Fano-type line shape.\cite{Ogita,Loosdrecht} 
For the infrared-active folded phonons, one B$_{3u}$, one B$_{1u}$ and two B$_{2u}$ modes were found at 98, 284, 312 and 800 cm$^{-1}$, respectively.\cite{Damascelli,Popova,Takehana5} 
The field dependence of the folded phonon modes was investigated in Raman experiments.\cite{Loosdrecht2,Loa2} 
The intensity decreases steeply at the boundary between the D and IC phases, while no energy shift is observed. 
Just above $H_C$, the intensity decreases to about half that of the D phase, and continues to decrease in the IC phase with increasing field. 
The folded mode at 312 cm$^{-1}$ was observed above $H_C$, 
but the details were unclear because it is located on the shoulder of an optical phonon.\cite{Musfeldt} 

In contrast to these results, in our previous paper,\cite{Takehana5} we have found a splitting of the folded phonon mode at 98 cm$^{-1}$ in the IC phase. 
The energy separation between the split-off branches is proportional to the incommensurability in the IC phase.
It indicates that the incommensurability is closely related to the mechanism of splitting of the folded phonon in the IC phase. 
In the IC phase, the phonon modes at ${\bf k} = \pm({\bf q}_{SP}- \Delta{\bf q})$ can be infrared active due to the incommensurate lattice modulation.\cite{Janssen} 
We have explained this phenomenon as follows: 
The 98 cm$^{-1}$ phonon is a spin-phonon coupled mode. 
Because of this the phonon mode at ${\bf k} = {\bf q}_{SP}- \Delta{\bf q}$ and that at ${\bf k} = -({\bf q}_{SP}- \Delta{\bf q})$ can interact with each other through the spin component of 2$\Delta{\bf q}$ of static incommensurate spin and lattice structure in the IC phase. 
The presence of 2$\Delta{\bf q}$ spin component was recently confirmed firmly by NMR\cite{Horvatic} and neutron diffraction\cite{Ronnow} experiments.

In CuGeO$_3$, soft mode has not been found yet and now it is believed that there is no soft mode. 
On the other hand, there should be spin-phonon coupled mode(s) in CuGeO$_3$, because there occurs spin-Peierls transition in it. 
Therefore, the finding of the spin-phonon coupled mode is very important for understanding various phenomena in CuGeO$_3$. 

In this paper, studies of the spin-phonon coupled modes near 98 cm$^{-1}$ are extended to doped CuGeO$_3$ and the behaviors of these modes will be clarified in the D, IC, DAF and IAF phases.

\section{Experimental}

Doped CuGeO$_{3}$ single crystals were grown by a floating zone method using an image furnace and were cleaved along the (100) plane.  
The sample of CuGe$_{0.988}$Si$_{0.012}$O$_{3}$ with dimensions of 1.5 $\times$ 4 $\times$ 6 mm$^{3}$ was used in this study. 
It showed two phase transitions, U-D and D-DAF, at low temperatures in zero field. 

Far-infrared (FIR) transmission was measured in the spectral range between 15 and 350 cm$^{-1}$ with a maximum resolution of 0.1 cm$^{-1}$ using a Fourier transform spectrometer (BOMEM DA8). 
The polarized measurements were performed in zero field by using FIR polarizers. 
The unpolarized magneto-optical spectra were obtained with a superconducting magnet in the Faraday configuration. 
The temperature dependence of the spectra was investigated down to 2 K.
The spectra on the increasing (decreasing) field process were taken with a fixed field for several hours after increasing (decreasing) applied magnetic filed from lower (higher) field region.

\section{Results}

\begin{figure}
\centerline{\includegraphics*[width=7.5cm]{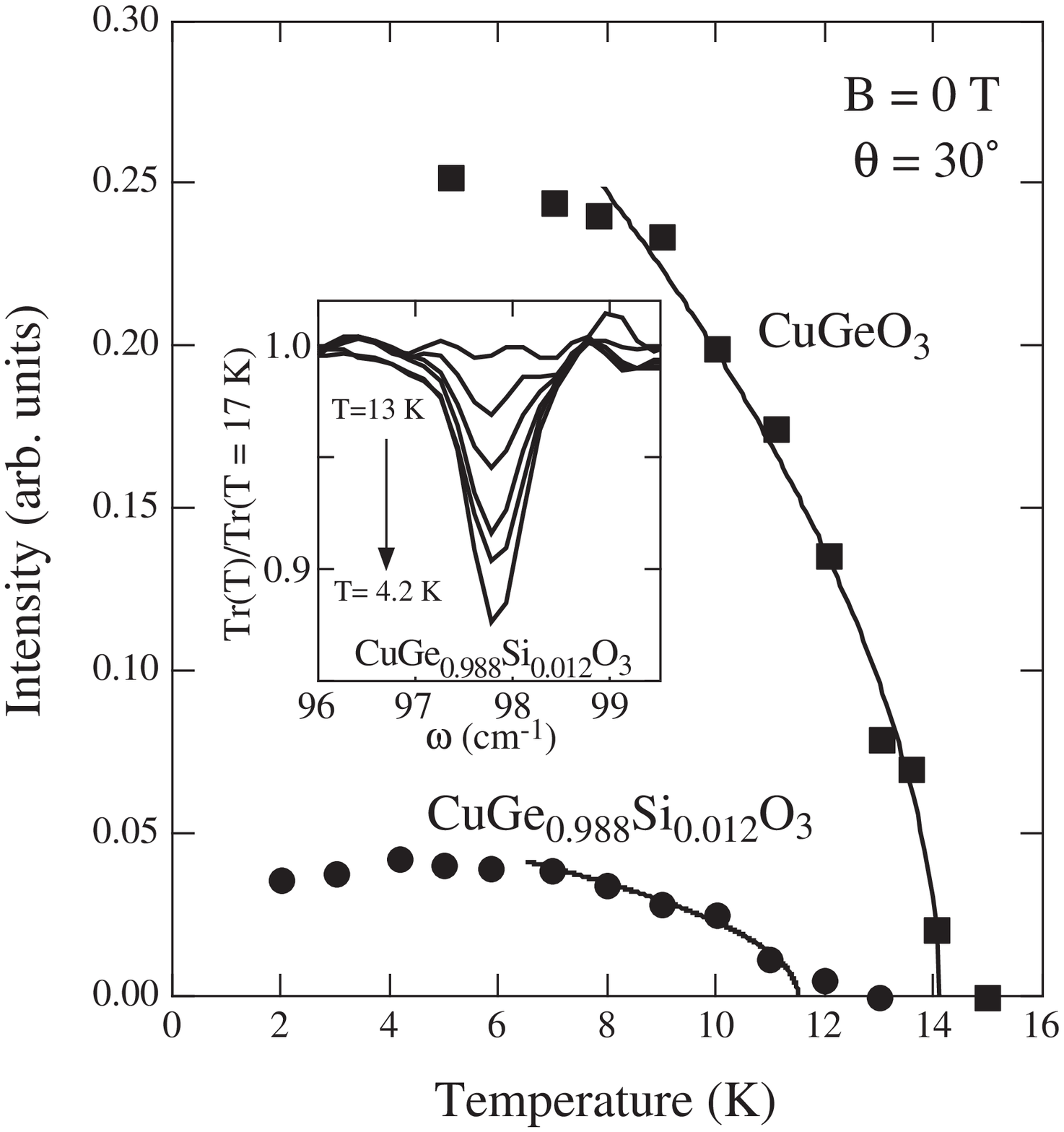}}
\caption{Temperature dependence of the intensity of FP on CuGe$_{0.988}$Si$_{0.012}$O$_{3}$ and CuGeO$_3$ in zero field with the samples rotated by  30$^{\circ}$ around the $b$ axis. 
The solid lines indicate the best fitted functions of $(1 - T/T_{SP})^{2\beta}$. 
The best fitted values of $2\beta$ are 0.55 and 0.5 for the pure and doped CuGeO$_3$, respectively. 
The inset shows that the absorption of FP increases with decreasing temperature down to $T_N \approx$ 3.5 K on CuGe$_{0.988}$Si$_{0.012}$O$_{3}$.}
\label{fig1}
\end{figure}

The doping effect of the folded mode at 98 cm$^{-1}$ was investigated on CuGe$_{0.988}$Si$_{0.012}$O$_{3}$ in the D, IC, DAF and IAF phases. 
The transmission spectra were normalized by the spectrum in the U phase, Tr({\em T} = 17 K, {\em B} = 0 T), in order to clarify the small change between the U and other phases. 
The inset of Fig.~\ref{fig1} shows the normalized transmission spectra, Tr($T$)/Tr($T$ = 17 K) at $B$ = 0 T on the doped sample. 
The spectra were taken when the sample was rotated around the $b$ axis by 30$^{\circ}$, so that the $a$ axis was 30$^{\circ}$ from the incident light direction (see Fig.~3 of Ref.~\onlinecite{Takehana5}), 
because an absorption line at 98 cm$^{-1}$ was assigned to the B$_{3u}$ folded phonon mode with the polarization property of ${\bf E} \parallel a$.\cite{Takehana5} 
The absorption appears at 98 cm$^{-1}$ below $T_{SP}$ and grows with decreasing temperature. 
Hereafter, the folded mode at 98 cm$^{-1}$ is labeled as FP. 
Figure~\ref{fig1} shows the absorption intensity of the FP of CuGe$_{0.988}$Si$_{0.012}$O$_{3}$ as a function of temperature, together with that of the pure CuGeO$_{3}$. 
The solid lines show the best fitted function $(1 - T/T_{SP})^{2\beta}$, where the best fitted value of $2\beta$ for the doped CuGeO$_3$ is 0.5, and that for the pure CuGeO$_3$ is 0.55.\cite{Takehana5} 
The dependence on the temperature around $T_{SP}$ is described well by the function $(1 - T/T_{SP})^{2\beta}$, 
while the SP transition of the doped CuGeO$_3$ seems to be somewhat less sharp and the FP was observed even just above $T_{SP}$, as was observed on other folded modes on the Raman spectra.\cite{Kuroe4} 
The intensity of the superlattice reflections of the doped CuGeO$_3$ exhibits a similar temperature dependence.\cite{Regnault,Sasago,Martin} 
The $T_{SP}$ of the doped sample shifts toward a lower temperature.\cite{Hase2} 
The FP mode was also observed below $T_N \approx$ 3.5 K with the same peak energy and linewidth as in the D phase. 
The absorption intensity of FP has a broad maximum around 4 K and slightly decreases with temperature in the DAF phase.

\begin{figure}
\centerline{\includegraphics*[width=7.5cm]{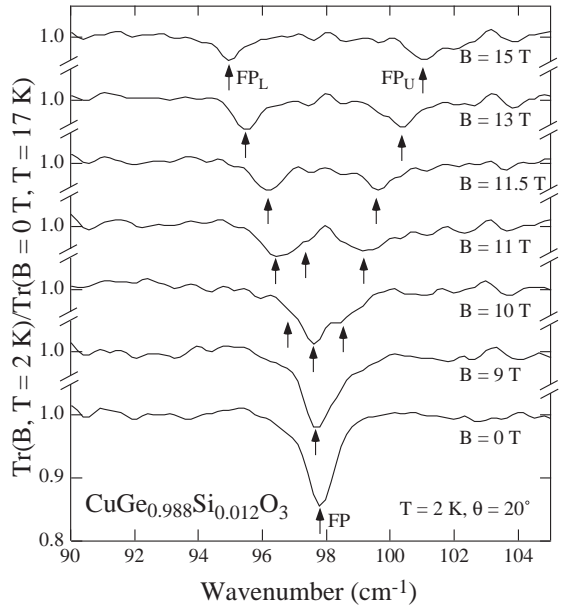}}
\caption{Field dependence of the 98 cm$^{-1}$ folded mode of CuGe$_{0.988}$Si$_{0.012}$O$_{3}$ at 2 K with the sample inclined around the $b$ axis by about 20$^{\circ}$ from the Faraday configuration ({\bf B} $\parallel a$).}
\label{fig2}
\end{figure}

Figure~\ref{fig2} shows the field dependence of FP mode of the doped CuGeO$_3$ at $T$ = 2 K on the decreasing field process with the sample rotated around the $b$ axis by about 20$^{\circ}$. 
The region below $H_C$ belongs to the DAF phase, and that above $H_C$ does to the IAF phase. 
The splitting of FP into two branches, which are labeled as FP$_U$ and FP$_L$ in Fig.~\ref{fig2}, were observed above $H_C$ on the doped CuGeO$_3$, 
as was observed in pure CuGeO$_3$ at the transition from the D phase to the IC phase.\cite{Takehana5} 
The energy separation $\Delta\omega$ between FP$_U$ and FP$_L$ increases with field.
The $H_C$ of the doped CuGeO$_3$ is lower than that of the pure one, as was observed in Ref.~\onlinecite{Hase4}. 
The coexistence of FP, FP$_U$ and FP$_L$ was observed around $H_C$ on the doped sample.
The intensities of FP$_U$ and FP$_L$ in the higher field region are almost equal to each other, and roughly a quarter of that of FP in zero field. 
The halfwidths of FP$_U$ and FP$_L$ are almost the same as that of FP, while it is difficult to estimate them around $H_C$ owing to the overlap of the lines.

\begin{figure}
\centerline{\includegraphics*[width=7.5cm]{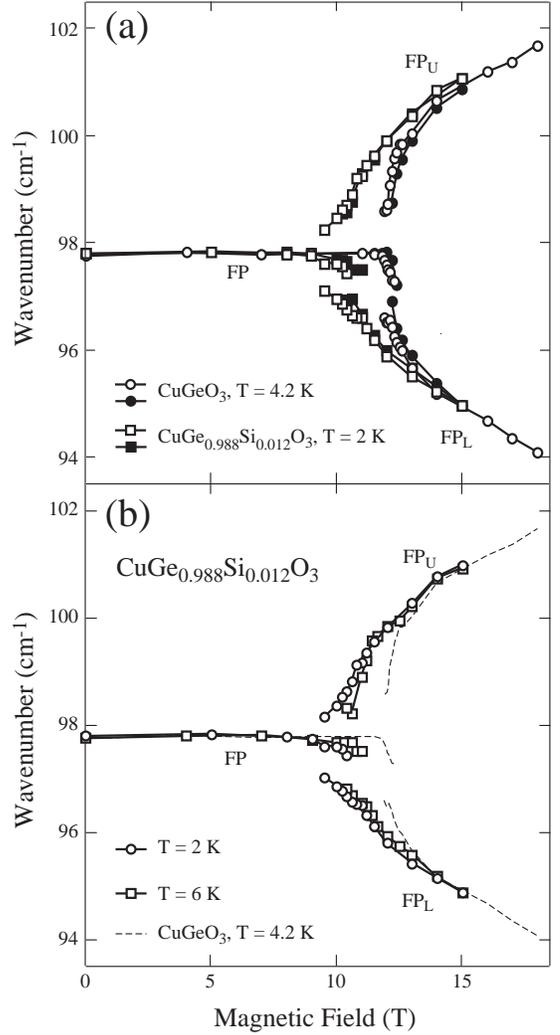}}
\caption{(a)Field dependence of the peak positions of FP, FP$_U$ and FP$_L$ at $T$ = 2 K on CuGe$_{0.988}$Si$_{0.012}$O$_{3}$ and that at $T$ = 4.2 K on CuGeO$_{3}$ (Ref.~\protect\onlinecite{comment1}), 
when the sample is rotated by  20$^{\circ}$ from the Faraday configuration ({\bf B} $\parallel a$). 
Open symbols show the data on the decreasing field process and closed symbols do those on the increasing field process. 
Hysteresis appears around the critical field, $H_C$. 
(b)Field dependence of the energy positions of FP, FP$_U$ and FP$_L$ at $T$ = 2 K and 6 K on CuGe$_{0.988}$Si$_{0.012}$O$_{3}$ on the decreasing field process. 
The broken lines show the field dependences of these peak energies on CuGeO$_{3}$ at $T$ = 4.2 K, for reference. }
\label{fig3}
\end{figure}

The peak energies of FP, FP$_U$ and FP$_L$ on CuGe$_{0.988}$Si$_{0.012}$O$_{3}$ are plotted in Fig.~\ref{fig3}(a) as functions of the applied magnetic field on both the increasing and decreasing field processes. 
For reference, the field dependences of CuGeO$_3$ are also plotted in Fig.~\ref{fig3}(a).
The energies of FP do not depend on the applied field below $H_C$. 
Shift of FP toward lower energies around $H_C$ was observed on the doped sample, but the amount of the shift is much smaller than that of the pure sample. 
The positions of FP$_U$ and FP$_L$ are symmetrical with respect to that of FP of $H\leq H_C$  on both the pure and doped CuGeO$_3$. 
$\Delta$$\omega$ increased with field above $H_C$, and the rate of change of $\Delta$$\omega$ with respect to the applied field decreases gradually with increasing field.
The difference of behaviors between the pure and doped CuGeO$_3$ was observed on the field dependence of $\Delta$$\omega$ around $H_C$, 
that is, the increasing rate of $\Delta$$\omega$ on the doped CuGeO$_3$ is relatively lower than that on the pure one. 
The field dependence of FP, FP$_U$ and FP$_L$ on CuGeO$_3$ exhibits a hysteresis around $H_C$. 
The presence of the hysteresis means that the transition between the D and IC phases is of the first order even in the doped sample 
and this corresponds to the fact that the hysteresis is also observed in pure CuGeO$_3$ in magnetization,\cite{Hase2} magnetostriction\cite{Takehana1} and the incommensurability of the lattice modulation.\cite{Kiryukhin1} 
The amounts of the hysteresis of FP$_U$ and FP$_L$ on both the pure and doped samples are about $\Delta$$B_H \approx$ 0.2 T in the region of $\Delta$$\omega \approx$ 3 cm$^{-1}$, which were estimated by the difference between the magnetic fields where $\Delta$$\omega$s are equal to each other on the increasing and decreasing field processes. 
They are almost independent of the field around $H_C$. 
In the doped sample, the remnant signal of FP on the increasing field process and those of FP$_U$ and FP$_L$ on the decreasing field process were observed in a relatively wide range of $\Delta$$B_R \approx$ 0.4 T. 
The remnant signals have not been observed in the pure CuGeO$_3$. 
The slow field dependence and the remnant signals of these branches make the coexistence region of the doped CuGeO$_3$ much wider than that of the pure one. 
Enhancement of the coexistence region on the doped CuGeO$_3$ was also observed by x-ray experiments.\cite{Kiryukhin3} 
Hereafter, we show only the field dependence of FP, FP$_U$ and FP$_L$ on the decreasing field process. 

Figure~\ref{fig3}(b) shows the field dependences of the peak energies 
of FP, FP$_U$ and FP$_L$ on CuGe$_{0.988}$Si$_{0.012}$O$_{3}$ at $T$ = 6 K 
together with those at $T$ =  2 K.
The broken lines show the field dependences of these peak energies on CuGeO$_{3}$ at $T$ = 4.2 K, for reference. 
The region near  6 K is well above $T_N$ at $B=0$ T and therefore at $B\sim 15$ T
it should belong to the U phase according to 
Ref.~\onlinecite{Poirier} or to the IC phase according to 
Refs.~\onlinecite{Hiroi} and \onlinecite{Takeya}. 
It shows definitely that the FP mode is split into two modes 
FP$_U$ and FP$_L$ above $H_C \sim$ 10.5 T.
This fact clearly shows that a phase with an incommensurate structure 
exists above $H_C$ at $T=$ 6~K, which we may call IC phase.
Therefore we have confirmed that the phase diagram obtained 
in Refs.~\onlinecite{Hiroi} and \onlinecite{Takeya} is realized in 
CuGe$_{0.988}$Si$_{0.012}$O$_{3}$.
The field dependence of FP$_U$ and FP$_L$ at $T$ = 6 K is slightly steeper than that  at $T$ = 2 K in 
the region of $\Delta$$\omega <$ 3 cm$^{-1}$. 
On the other hand it overlaps with that at $T$ = 2 K 
in the region of $\Delta$$\omega >$ 3 cm$^{-1}$.
The remnant signals were also observed at $T$ = 6 K but 
the field range is much narrower than that at $T$ = 2 K.

\section{Discussion}

In this study, the doping effect of the folded mode FP was investigated in the D, IC, DAF and IAF phases on the doped CuGeO$_{3}$. 
The folded mode FP appears on the doped sample below $T_{SP}$, and the peak energy is independent of the doping in the D phase. 
The intensity of the doped sample is much weaker than that of the pure one. 
A similar behavior was seen in the absorption intensity of the 800 cm$^{-1}$ folded phonon mode.\cite{Damascelli2} 
The halfwidth of the FP mode on the CuGe$_{0.988}$Si$_{0.012}$O$_{3}$ is about 1.5 times wider than that on the pure CuGeO$_{3}$. 
There are few reports on the effects of doping on the halfwidth of other folded phonon modes. 
That of the Raman-active 369 cm$^{-1}$ mode is unclear due to the low resolution of the measuring system.\cite{Kuroe3}
That of the infrared-active 800 cm$^{-1}$ mode was not estimated, although the halfwidth seems to be somewhat broadened in Fig.~10 of Ref.~\onlinecite{Damascelli2}. 

The FP mode was observed below $T_N$, because of the presence of the lattice modulation of ${\bf q}_{SP}$.\cite{Regnault,Sasago,Martin}  
Figure~\ref{fig1} shows that the intensity of the doped sample 
decreases below $T_N \approx$ 3.5 K with temperature.
This means that the dimerization is suppressed with the growth of 
the antiferromagnetic long-range order. 
This indicates that the dimerization (SP order parameter) coexists with the antifferomagnetic long-range order in the DAF phase and that these two order parameters interact with each other. 
In the neutron diffraction measurements, the superlattice reflection peak was found to coexist with the antiferromagnetic peak and the intensity of the superlattice reflection decreases.\cite{Regnault,Sasago,Martin} 
The temperature dependence of the intensity of FP is fully consistent with that of the superlattice reflection because the intensity of the folded phonons is closely related to the magnitude of the lattice distortion caused by the dimerization due to the SP ordering. 

Above $H_C$, the splitting into FP$_U$ and FP$_L$ was observed in the IC and IAF phases on the doped sample. 
Based on our model\cite{Takehana5} of the splitting of the folded phonon FP, the necessary conditions for splitting are not only the spin-phonon coupling 
but also the existence of the lattice modulation of $\pm({\bf q}_{SP} - \Delta{\bf q})$ and the component of the spin polarization with the periodicity of 2$\Delta{\bf q}$. 
Therefore, the experimental results indicate the existence of the  incommensurate modulation in the lattice and the spin polarization in both the IC and IAF phases. 
The incommensurate lattice modulation was observed in the IC and IAF phases on doped samples in the x-ray diffraction experiments.\cite{Kiryukhin3} 
The behavior of the splitting different from the pure CuGeO$_3$ was found around $H_C$. 
That is, the field dependence of the peak positions of FP$_U$ and FP$_L$ on the doped CuGeO$_3$ are relatively slower than that on the pure one. 
Moreover, the field dependence at $T$ = 2 K is slightly weaker than that at $T$ = 6 K in the vicinity of $H_C$, while those at $T$ = 2 K and 6 K almost overlap with each other in higher field region. 
The positions of FP$_U$ and FP$_L$ in the IC and IAF phases of the doped sample gradually approach to those of the pure one with increasing field, which means that their field dependences can be well described by the same function at higher field range. 

In our previous paper, it was shown that the field dependence of $\Delta$$\omega$ can be scaled with that of the incommensurability, $\Delta L$, and we concluded that $\Delta$$\omega$ is proportional to the $\Delta L$ in the pure CuGeO$_3$.\cite{Takehana5} 
In order to clarify the difference between the pure and doped CuGeO$_3$, $\Delta$$\omega$s of the doped CuGeO$_3$ in the IAF and IC phases are plotted by scaling them with $H_C$, as well as $\Delta$$\omega$ of the pure CuGeO$_3$ and $\Delta L$ in the x-ray experiments, as shown in Fig.~\ref{fig4}. 
The critical fields $H_C$ of the doped CuGeO$_3$ are determined to fit the field dependence of $\Delta$$\omega$ into the fitting function, especially in the higher field range where the similarity to the pure CuGeO$_3$ was observed, 
because it is difficult to define correct $H_C$ of the doped CuGeO$_3$ owing to the wide coexistent region. 
The scale of the right-side ordinate for $\Delta$$\omega$ is the same as that of the pure CuGeO$_3$, which is adjusted to fit the field dependence of $\Delta$$\omega$ to that of $\Delta L$. 
The solid curve in Fig.~\ref{fig4} is the fitting function $1/\ln[8/(H/H_C - 1)]$.\cite{Buzdin} 
The dashed line indicates the theoretical prediction by Cross for the high magnetic field limit.\cite{Cross} 
Figure~\ref{fig4} indicates that $\Delta$$\omega$ of the doped CuGeO$_3$ deviates from the fitting function in the IAF and IC phases in the lower field range of $H/H_C <$ 1.03, 
and are observed even at $H/H_C <$ 1, while they can be plotted quite well on the same universal curve as that of the pure one at higher field region of $H/H_C >$ 1.03. 
The discrepancy increases with decreasing $\Delta$$\omega$ around $H_C$. 
A similar behavior can be seen on $\Delta L$ of the doped CuGeO$_3$, especially, with higher concentration of impurities,\cite{Kiryukhin2} as shown in Fig.~\ref{fig4}. 
In the IAF phase, the discrepancy is larger than that in the IC phase on the doped CuGeO$_3$, and $\Delta$$\omega$  in the IAF survives at the lower field in $H/H_C$ scale. 
This result indicates that $\Delta$$\omega$ and $\Delta L$ are no longer described by the fitting function at this field range. 
This does not necessarily mean that the relation $\Delta\omega\propto\Delta L$ is broken in this region because $\Delta\omega$ and $\Delta L$ were measured on different samples.
In higher field region of $H/H_C >$ 1.03, $\Delta L$ is well described by the same theoretical fitting function on both pure and doped CuGeO$_3$. 
Therefore, the proportional relation of $\Delta$$\omega \propto \Delta L$ may be valid even on the doped CuGeO$_3$ at $H/H_C >$ 1.03. 
We assumed that the strength of the coupling is proportional to the density of the solitons with periodicity 2$\Delta{\bf q}$, i.e., the integrated intensity of the average component of the spin polarization of 2$\Delta{\bf q}$.\cite{Takehana5} 
Consequently, $\Delta$$\omega$ is determined only by the incommensurability, $\Delta L$, and is independent of the species of the impurities or the mechanism of how the local magnetization is induced around the impurities.\cite{Fukuyama,Hansen} 
$\Delta$$\omega$ approaches gradually the curve predicted by the theory of Cross with increasing field, and seems to approach the theory of Cross above $H/H_C \approx 1.5$. 
In the field range of $H/H_C \geq 1.5$,  the field dependence of the magnetization\cite{Hase2,Horvatic} and the spontaneous strain\cite{Lorenz,Takehana2} can be fitted by a linear function and the lattice modulation is well described as sinusoidal.\cite{Lorenz} 
The discrepancy between the pure and doped CuGeO$_3$ around $H_C$ can be explained by the presence of interaction between the solitons and the impurities, as follows. 
The interaction between the solitons and the impurities must exist on the doped CuGeO$_3$, and binding of the solitons on the impurities can occur.\cite{Hansen,Augier} 
They change the distribution of intersoliton distance, and induces randomness in it. 
The effect of the interaction is enhanced on the condition that the number of solitons is of the same order as that of the impurities, and is weakened with increasing soliton numbers as compared with the impurity numbers. 
Therefore, in higher field region, the effect of this interaction is weak and $\Delta$$\omega$ and $\Delta L$ exhibit almost the same field dependence as that of the pure CuGeO$_3$. 
On the other hand, around $H_C$, the effect of the interaction is no longer negligible, and the randomness prevents $H_C$ from being well-defined, which makes the field dependence of $\Delta$$\omega$ weaker. 
The different behavior of $\Delta$$\omega$ between the IAF and IC phases indicates that the impurity effect in the IAF phase is stronger than that in the IC phase. 
This is consistent with the model of the ^^ ^^ freezing" of the solitons in the IAF phase.\cite{Takeya} 
The observation of the remnant signals in the IAF phase would be also related with the ^^ ^^ freezing" of the solitons. 
Interestingly, the interaction between the solitons and the impurities, or ^^ ^^ freezing" of the solitons, do not affect the hysteresis of FP$_L$
and FP$_U$. 
The reduction of the shift of FP in the vicinity of $H_C$ might be connected with this interaction if, as we believe, the shift of FP is caused by the discommensuration which appears just above $H_C$.

\section{Conclusions}

We have performed the doping effect of the folded phonon mode at 98 cm$^{-1}$ on the Si-doped CuGeO$_3$. 
The folded phonon mode was observed in both the D and DAF phases, and the splitting of the mode was observed in the IC and the IAF phases. 
This definitely proves that IC phase exists in the region of $T_N \leq T \leq T_{SP}$ and $H \geq H_C$, which is consistent with the reports of Refs.~\onlinecite{Hiroi} and \onlinecite{Takeya}. 
The different behavior of the split-off branches from the pure CuGeO$_{3}$ was observed in the vicinity of $H_C$, and the discrepancy in the IAF phase is larger than that in the IC phase. 
It is explained by the interaction between the solitons and the impurities.

\begin{figure}
\centerline{\includegraphics*[width=7.5cm]{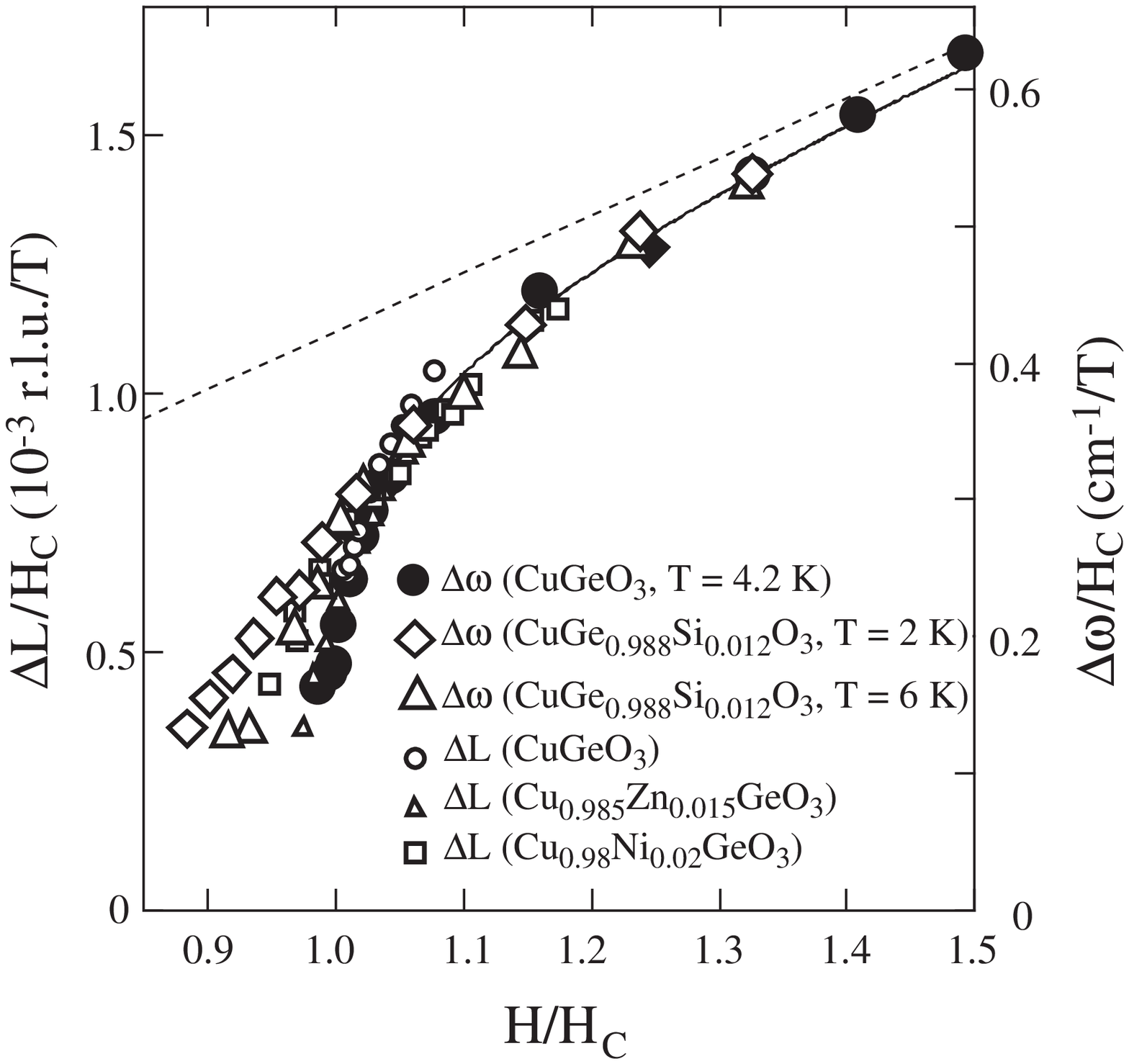}}
\caption{
Scaled Field dependence of the energy separation $\Delta$$\omega$ between the two split branches at $T$ = 2 K and 6 K on CuGe$_{0.988}$Si$_{0.012}$O$_{3}$ and that at $T$ = 4.2 K on CuGeO$_{3}$ (Ref.~\protect\onlinecite{comment1}). 
That of the superlattice reflections, $\Delta L$, on both the pure and doped CuGeO$_3$ are also plotted simultaneously, where $\Delta L$ were measured by Kiryukhin {\it et al.} (Ref.~\protect\onlinecite{Kiryukhin2}). 
Both $\Delta$$\omega$ and $\Delta L$ were scaled by the respective critical field, $H_C$. 
A dashed line indicates the theoretical prediction by Cross for the high magnetic field limit (Ref.~\protect\onlinecite{Cross}). 
A solid curve is the fitting function $1/\ln[8/(H/H_C - 1)]$, which was predicted by mean field theory (Ref.~\protect\onlinecite{Buzdin}).
}
\label{fig4}
\end{figure}

\end{multicols}

\begin{references}
\bibitem{comment1} $\Delta$$\omega$ of the pure CuGeO$_3$ in Fig.~\ref{fig4} was reexamined in this study because the hysteresis was not taken into account of in Ref.~\onlinecite{Takehana5}.
\end{references}
\end{document}